\documentclass[superscriptaddress, amsmath, amssymb, aps, prx, nofootinbib, twocolumn]{revtex4-2}

\usepackage[utf8]{inputenc} 
\usepackage[T1]{fontenc}    
\usepackage{hyperref}       
\usepackage{url}            
\usepackage{amsfonts}       
\usepackage[usenames]{xcolor}         
\usepackage{graphicx} %
\usepackage{amsmath}
\usepackage{lmodern}
\usepackage{amssymb}
\setcitestyle{numbers}
\usepackage[capitalize]{cleveref}
\usepackage[shortlabels]{enumitem}

\newcommand{\loss}{\mathcal{L}}

\newcommand{\diff}{\mathop{}\!\mathrm{d}}

\newcommand{\avg}[1]{\left\langle #1 \right\rangle}

\hypersetup{
	colorlinks,
    linkcolor={blue!80!black},
    citecolor={blue!80!black},
    urlcolor={blue!80!black}
	}
	
\begin{document}

\title{Charting the Topography of the Neural Network Landscape with Thermal-Like Noise}

\author{Th\'eo Jules}
\email{theo.jules.physics@gmail.com}
\affiliation{Raymond and Beverly Sackler School of Physics and Astronomy, Tel Aviv University, Ramat Aviv, Tel Aviv, 69978, Israel}
\author{Gal Brener}
\affiliation{Raymond and Beverly Sackler School of Physics and Astronomy, Tel Aviv University, Ramat Aviv, Tel Aviv, 69978, Israel}
\author{Tal Kachman}
\affiliation{Department of Artificial Intelligence, Radboud University, Thomas van Aquinostraat 4, 6525 GD Nijmegen}
\author{Noam Levi}
\affiliation{Raymond and Beverly Sackler School of Physics and Astronomy, Tel Aviv University, Ramat Aviv, Tel Aviv, 69978, Israel}
\author{Yohai Bar-Sinai}
\email{ybarsinai@gmail.com}
\affiliation{Raymond and Beverly Sackler School of Physics and Astronomy, Tel Aviv University, Ramat Aviv, Tel Aviv, 69978, Israel}
\affiliation{The Center for Physics and Chemistry of Living Systems, Tel Aviv University, Tel Aviv 69978, Israel}


\begin{abstract}
The training of neural networks is a complex, high-dimensional, non-convex and noisy optimization problem whose theoretical understanding is interesting both from an applicative perspective and for fundamental reasons. A core challenge is to understand the geometry and topography of the landscape that guides the optimization. In this work, we employ standard Statistical Mechanics methods, namely, phase-space exploration using Langevin dynamics, to study this landscape for an over-parameterized fully connected network performing a classification task on random data. Analyzing the fluctuation statistics, in analogy to thermal dynamics at a constant temperature, we infer a clear geometric description of the low-loss region. We find that it is a low-dimensional manifold whose dimension can be readily obtained from the fluctuations. Furthermore, this dimension is controlled by the number of data points that reside near the classification decision boundary. Importantly, we find that a quadratic approximation of the loss near the minimum is fundamentally inadequate due to the exponential nature of the decision boundary and the flatness of the low-loss region. This causes the dynamics to sample regions with higher curvature at higher temperatures, while producing quadratic-like statistics at any given temperature. We explain this behavior by a simplified loss model which is analytically tractable and reproduces the observed fluctuation statistics.
\end{abstract}

 \maketitle
The optimization of neural networks lies at the core of modern learning methodology, with the goal of minimizing a loss function that quantifies model performance. Naturally, the landscape of the loss function plays a critical role in guiding the optimization process and its properties are closely linked to its performance and generalization capacities~\cite{Dinh2017, Yang2021}. However, the high dimensionality of the parameter space, the non-convexity of the loss function, and the presence of various sources of noise make it challenging to characterize its geometry~\cite{Li2018a, czarnecki2019deep} and subsequently to analyze the optimization process over this complicated landscape.

Previous works have studied the topography of the loss landscape and found a number of interesting features. Firstly, it was established that there exists a wealth of global minima, all connected by low-loss paths, a phenomenon referred to as Linear Mode Connectivity~\cite{Draxler2018, Garipov2018, Fort2019, Fort2019b, Nguyen2019, Cooper2021, Benton2021}. In the final stages of training the network explores this low-loss region and gradient descent predominantly occurs within a small subspace of weight space~\cite{GurAri2018, Li2018, Frankle2020}. In addition, it was seen that the curvature of the explored region sharpens progressively and depends on the learning rate through a feedback mechanism termed ``Edge of Stability''~\cite{Cohen2021, Wang2022, Damian2022}.

In this work we study the low loss region by injecting noise in a controlled manner during training. Many previous works have studied the importance of noise in the optimization process, modeling it as a stochastic process. Noise sources might include sampling noise in the estimation of the gradient~\cite{Zhu2018,Simsekli2019,Xie2020}, the numerical discretization of gradient flow~\cite{Elkabetz2021}, noisy data~\cite{Levi2022, Levi2022a}, stochastic regularization schemes~\cite{Srivastava2014} or other sources. Each such noise source gives rise to different noise properties, which qualitatively affect the optimization dynamics~\cite{Li2021, Li2021a}. 

We take a different approach than those described above: we do not use noise to mimic noisy training dynamics, but rather as a probe that allows inferring quantitative geometrical insights about the loss landscape~\cite{Levi2022, Levi2022a}. This is done using standard tools of statistical physics to analyze loss fluctuations, and ensuring that the thermal noise is the only noise source in the system so the stochasticity is completely known.

To study the local landscape, we let the system evolve, starting at the minimum, under over-damped Langevin dynamics, defined by the stochastic differential equation 
\begin{align}
\diff \theta_t &= -  \nabla_\theta \loss(\theta_t) \diff t + \sqrt{2T} \diff  W_t,
\label{eq:SDE_Langevin}
\end{align}
where $\theta\in\mathbb{R}^N$ is the vector of the neural weights and biases, $\loss$ is the loss function (to be specified below), $T$ the exploration temperature and $W_t$ is a standard $N$-dimensional Wiener process. In terms of statistical physics, this is analogous to a system whose phase space coordinates are $ \theta$ and which is described by a Hamiltonian $\loss(\theta)$ in contact with a thermal bath at temperature $T$. As is well known~\cite{Tuckerman2010}, the long time limit of the probability distribution of $\theta$ is a Boltzmann distribution, $p(\theta)\propto e^{-\loss(\theta)/T}$, which balances between the gradient and the random noise terms in Eq.~\eqref{eq:SDE_Langevin}.

Specifically, we explore the topography of the loss function in the vicinity of a typical minimum, for a simple fully connected network performing a classification task of random data in the over-parameterized regime. 
Our analysis shows that, for the networks that we studied, the minimum is constrained only in a small number of directions in weight-space, as was previously observed in various contexts and is generally expected in the over-parameterized regime~\cite{Fort2019, Brea2019, Yang2021, Simsek2021, Ainsworth2022, GurAri2018, Li2018, Frankle2020}. Furthermore, and inline with previous studies, we find that at a given exploration temperature the fluctuations behave as if $\loss$ is effectively quadratic, with $N_c$ independent degrees of freedom with non-vanishing stiffness. In other words, $N_c$ is the co-dimension of the low-loss manifold in the vicinity of the minimum, which our method allows to measure directly. 

However, contrary to previous works and quite counter-intuitively, we show that this picture \textit{does not} arise from a simple quadratic approximation of $\loss$ around its minimum, as one might na\"{i}vely interpret these observations. Instead, we find that the stiffness associated with the $N_c$ constrained eigendirections depends linearly on $T$ over many orders of magnitude, which is a distinctly nonlinear feature. As we explain below, this dependence stems from the exponential nature of the ``confining walls'' surrounding the low-loss region, and the flatness of the landscape far from these walls. This exponential nature is also what gives rise to the seemingly quadratic properties of the loss fluctuations, but this happens through a delicate balance between the exponential walls and the noise, which cannot be captured with a model of a quadratic loss function. 

\section{Exact predictions for a quadratic loss}
Before describing our results, it would be useful to remind the reader what they would expect to observe in the case of a positive-definite quadratic loss function,
  $\loss=\sum_{i=1}^{N_c} \frac{1}{2}k_i \Theta_i^2\ $,
where $\{\Theta_i\}$ are the coefficients of the Hessian's eigenvectors and $\{k_i\}$ are their associated stiffnesses. $N_c$ is the number of dimensions with non-vanishing stiffness.
Plugging this into Eq.~\eqref{eq:SDE_Langevin} yields a multivariate Ornstein-Uhlenbeck process which is fully tractable analytically~\cite{Gardiner1985}. We briefly summarize here the main results, whose derivations can be found in the supplementary information. 

First, the fluctuations of $\loss$ follow a $\Gamma$-distribution
\begin{align}
  P(\loss; \alpha, \beta) &= \frac{\beta^\alpha \loss^{\alpha-1}}{\Gamma(\alpha)}\exp(-\beta\loss),
\label{eq:gamma_pdf}
\end{align}
where $\alpha = N_c/2$, $\beta=1/T$ and $\Gamma$ is the Gamma function. 

Second, a direct corollary of Eq.~\eqref{eq:gamma_pdf} is that the mean and standard deviation of $\loss$ are both proportional to $T$:
\begin{align}
\begin{split}
\mu_\loss&= \avg{\loss} =\tfrac{1}{2}N_cT,\\
\sigma_\loss^2 &= \avg{\loss^2}-\avg{\loss}^2=\tfrac{1}{2}N_cT^2\ .
\end{split}
\label{eq:mu_sigma_quadratic}
\end{align}
This result, a standard example of the equipartition theorem~\cite{Kardar2007}, means that each eigendirection contributes $\frac{1}{2}T$ to the total loss, regardless of its associated stiffness. The ``heat capacity'' $C_h=\partial\loss/\partial T$ simply equals $N_c/2$ and is $T$-independent.

Lastly, in terms of dynamics, the evolution of each eigendirection is uncorrelated from the other ones and shows an exponentially decaying correlation. This is quantified by the two-point correlation
 \begin{align}
  \chi_{g}(t)
  &= \sigma_{g}^{-2}\left[\avg{\vphantom{\Big[}g(t_0)g(t_0+ t)}-\mu_{g}^2\right]
  \label{eq:Auto-correlation}
\end{align}
where $g$ is any time-dependent quantity. For a quadratic loss we have $\chi_{\Theta_i}=\exp\left(-|t|/\tau_i\right)$ and the correlation time $\tau_i$ is simply the inverse of the stiffness $\tau_i=1/k_i$. We note that in these terms, the stiffness of the ``soft directions'' does not need to strictly vanish -- $k_i$ should only be low enough so that the correlation time $\tau_i$ would be so long that the dynamics in this eigendirection would not equilibrate during the simulation time.  The auto-correlation of $\loss$ is a sum of such exponentials, $\chi_\loss=\sum_i e^{-k_i|t|}$.

\section{Numerical experiment}
We consider a classification problem with $C=3$ classes using a multi-layer perceptron~\cite{Murphy2012}, represented by the function $f(x;\theta)\!:\!\mathbb R^{d}\to \mathbb R^C$. 
The network is trained on a training dataset $\{x^i, y^i\}_{i=1}^D$ where $x^i \in \mathbb R^{d}$ are the inputs and $y^i\in \{0,1\}^C$ are one-hot vectors indicating a randomly assigned correct class. The $\{x_i\}$  are drawn from a standard $d$-dimensional normal distribution. Full details regarding the architecture of the network and the dataset are given in the supplementary information.
The network's output is transformed to a classification prediction via a softmax function. That is, the estimated probability that an input $x^i$ belongs to class $k$ is
\begin{align}
  p_k(x^i;\theta) = \frac{\exp(f(x^i;\theta)_k)}{\sum_{m=1}^{C} \exp(f(x^i;\theta)_m)}\ ,
  \label{eq:prob_out}
\end{align}
where $f(\cdot)_k$ denotes the $k$-th entry in $f$.
Finally, the loss is taken to be the cross entropy between the predicted and true labels:
\begin{align}
  \loss &= \frac{1}{D} \sum_{i=1}^D \ell(x^i, y^i, \theta)\ , &
  \ell &= - \sum_{k=1}^C y^i_k \log(p_k(x^i;\theta)).
  \label{eq:CE_loss}
\end{align}

Our main objective is to explore the topography of the loss function in the vicinity of a typical minimum. To find such a minimum, we train the network using the ADAM optimizer~\citep{Kingma2014} for a predefined amount of epochs.
Since the problem is over-parameterized, after some training, the data is perfectly fitted and the loss becomes essentially zero, up to numerical noise. This stage is denoted as ``Adam'' in Fig.~\ref{fig:fig_1}a.

To explore the vicinity of this minimum, we then let the system evolve under Eq.~\eqref{eq:SDE_Langevin} using the Euler-Maruyama discretization scheme~\cite{higham2001algorithmic},
\begin{align}
\theta_{s+1} &= \theta_{s} - \eta \nabla \loss(\theta_{s}) + \sqrt{2\eta T} \xi_s \ ,
\label{eq:ULA}
\end{align}
where $s$ is the step number, $\eta=t_{s+1}-t_s$ is the discrete time step and $\xi_t$ is a Gaussian random variable with zero mean and unit variance. This exploration is denoted as ``Langevin'' in Fig.~\ref{fig:fig_1}. It is seen that the loss increases quickly before reaches a $T$-dependent steady state (``thermodynamic equilibrium''). 

We stress while that the parameter $\eta$ is reminiscent of the ``learning rate'' in the machine learning literature, they are not exactly equivalent. Importantly, in our formalism $\eta$ serves only as the time discretization and appears explicitly in the noise term, whose $\sqrt \eta$ scaling is necessary in order for the dynamics to converge to the Boltzmann distribution in the limit $\eta\to0$~\cite{Tuckerman2010}. We also note that the convergence of the probability distribution $p(\theta)$ in the limit $\eta\to0$ is a different concept than the convergence of the gradient descent trajectory to that of gradient flow~\cite{Elkabetz2021}. As such, $\eta$ is not a parameter of our exploration protocol but rather of the numerical implementation of Eq.~\eqref{eq:SDE_Langevin}, and meaningful results should not depend on $\eta$. 

\section{Results: Loss fluctuation statistics}

\begin{figure}[tb]
  \centering
  \includegraphics[width=\linewidth]{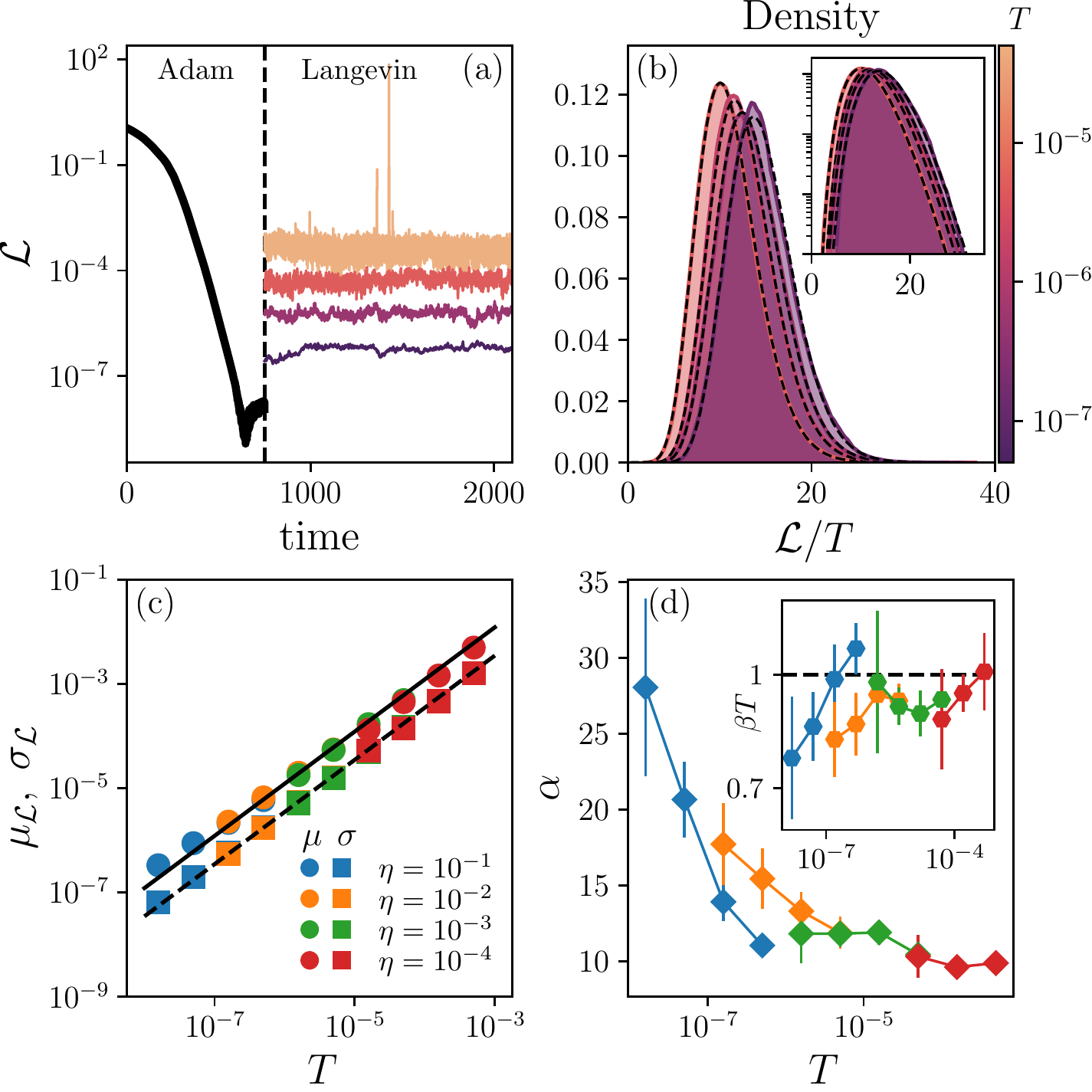}
  \caption{
  (a) Observed loss dynamics during the exploration. 
  First, the network is trained using the ADAM algorithm (black line). 
  Then, the learning algorithm is changed to Eq.~\eqref{eq:SDE_Langevin}, where the noise amplitude is controlled by a temperature-like parameter $T$ (colored lines). Each curve corresponds to a different temperature, all using $\eta = 10^{-2}$.
  (b) Distribution of the loss fluctuation in steady state normalized by the temperature. For each distribution, the dashed black line corresponds to a gamma distribution, cf~\cref{eq:gamma_pdf}, whose parameters are found using maximum likelihood estimation. The inset shows the same data in log-linear axes.
  (c) Temperature dependence of $\mu_\loss$ (circles) and $\sigma_\loss$ (squares). Each point corresponds to an average over multiple runs. The solid line shows a fit to $\mu_\loss = C_hT$. The dashed line shows the equilibrium prediction $\sigma_\loss=\sqrt{C_h} T$ with the obtained value of $C_h$.
  (d) Corresponding parameters $\alpha$ and $\beta$ for the gamma distribution.
  The symbols and error bars show the average and standard deviation, respectively, over multiple runs.}
  \label{fig:fig_1}
\end{figure}

We begin by inspecting the moments of the loss fluctuations, $\mu_\loss$ and $\sigma_\loss$, shown in Fig.~\ref{fig:fig_1}c. It is seen that both of them scale linearly with $T$. First, we note that our measurements of $\mu_\loss$ at a given temperature are independent of $\eta$, as expected. Furthermore, a basic prediction of statistical mechanics relates the variance of $\loss$ in equilibrium with the heat capacity, namely $\sigma_\loss^2=T^2 C_h(T)$~\cite{Kardar2007}. In our case of a $T$-independent heat capacity this relation reads $\sigma_\loss=\sqrt C_h T$, which is numerically verified in Fig.~\ref{fig:fig_1}c. These results support our claim that the dynamics are thermally equilibrated and follow Boltzmann statistics.

Going beyond the moments, Fig.~\ref{fig:fig_1}b shows the full distribution of the loss fluctuation, which are well described by a Gamma distribution. Fig.~\ref{fig:fig_1}d shows the distribution parameters $\alpha$ and $\beta$, defined in Eq.~\eqref{eq:gamma_pdf}, which are estimated from the empirical loss distributions using standard maximum likelihood estimators. It is seen that the distribution parameter $\beta$ agrees with the exploration temperature $T$, i.e.~$\beta T\approx 1$, over several orders of magnitude in $T$ and independently of $\eta$. The number of stiff dimensions, $N_c=2\alpha$, seems to weakly depend on the temperature, decreasing as $T$ grows. Lastly, we note that that the linear dependence of $\mu_\loss$ and $\sigma_\loss$ on $T$ is a property of the low-loss region explored by the dynamics at low $T$, and it is not observed if the thermal dynamics are started immediately after initializing the network. This is shown explicitly in the supplementary information.

All these observations are \emph{quantitatively} consistent with a picture of a (locally) quadratic loss function. In other words, at each temperature we can interpret the loss statistics as if they were generated by an effective quadratic loss, which has a $T$-dependent number of stiff directions, $N_c(T)=2\alpha(T)$. This number, $N_c\approx 20-60$, is significantly lower than both the dimensionality of $\theta$ ($N=900$) and the number of elements in the dataset, $D=300$. It is also much \emph{larger} than the number of classes $C=3$, which was suggested by Fort et.~al.~\cite{Fort2019} as the number of outlying large Hessian eigenvalues.

We find that the effective dimension of the low loss manifold is directly related to the number of points that lie close to the decision boundary. To demonstrate this, we examine the loss $\loss$ of \cref{eq:CE_loss} as a sum over the losses of individual sample points $\loss=D^{-1}\sum_i \ell_i$. We find numerically that most of the sample points are well classified, contributing negligibly to the total loss. A common way to quantify how many points contribute non-negligibly is the ratio of the $L_1$ and $L_2$ norms of the loss vector~\cite{Hurley2009},
\begin{align}
  \phi\left(\{\ell_i\}\right) = \frac{\left(\sum_{i=1}^D\ell_i\right)^2}{\sum_{i=1}^D\ell_i^2} \ ,
  \label{eq:participation_ratio}
\end{align}
where $\ell_i$ is the contribution of the $i$-th example to the loss. $\phi$ is a measure of sparsity, which counts how many entries in $\ell_i$ contribute to its sum $d$. For instance, if $\ell_1=\ell_2=\cdots=\ell_k$ and all other $\ell_i$ vanish then $\phi=k$. We calculate $\phi$ for random snapshots of the network during the dynamics, and plot the averaged results in \cref{fig:loss_autocorrelation}a. It is seen that $\phi$, the effective number of sample points pinning the decision boundary, quantitatively agrees with $\alpha$, twice the effective number of constrained dimension in weight space.

\subsection{Temperature dependence} 
\begin{figure}[tb]
  \centering
  \includegraphics[width=\linewidth]{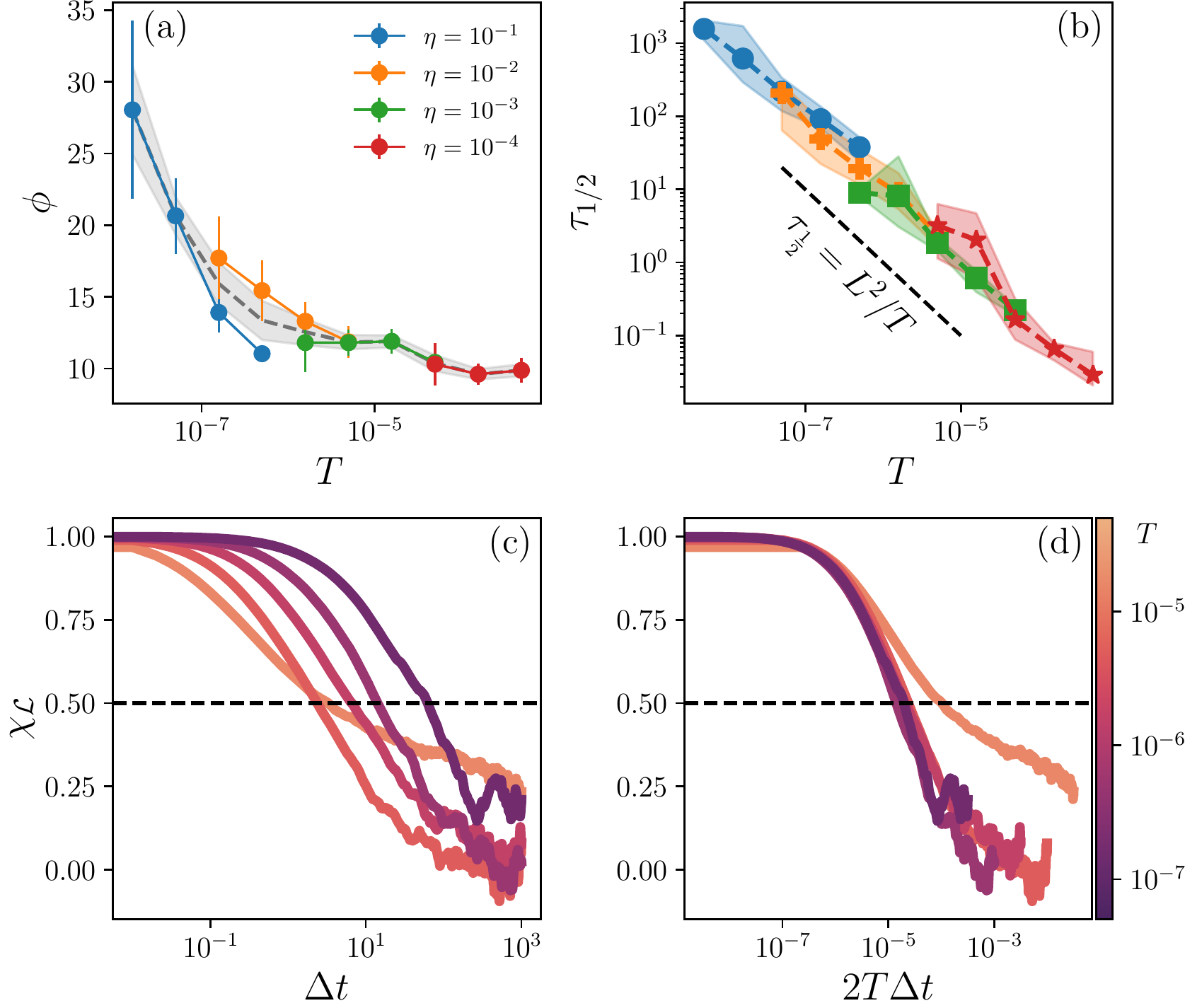}
  \caption{
    (a) The sparsity $\phi$, cf.~\cref{eq:participation_ratio}, as a function of $T$. In gray we overlay our estimations of $\alpha$, plotted in \cref{fig:fig_1}. It is seen that $\phi$ quantitatively agrees with $\alpha$, twice the effective number of constrained dimensions of the low loss manifold.
    (b) Temperature dependence of  $\tau_{\frac{1}{2}}$.
    The measurement was repeated over multiple runs, and the plot shows the average (points) and maximum and minimum values (color shading). 
    The black line shows a power law dependence $\tau_{\frac{1}{2}} = \frac{L^2}{T}$.
    (c) Autocorrelation of the loss (cf.~Eq.~\eqref{eq:Auto-correlation}) in steady-state.
    The correlation half-time $\tau_{\frac{1}{2}}$ is the time for which $\chi_\loss=0.5$.
    It is seen that the auto-correlation decays logarithmically at large $\Delta t$. 
    (d) The same data as in panel c, plotted as a function of the rescaled time lag $2T \Delta t$. The curves for different $T$ collapse to a single curve, except at high temperature and long times.
    }
  \label{fig:loss_autocorrelation}
\end{figure}

However, while the time-independent statistics suggest an effective quadratic loss, the dynamic properties show that the picture is not as simple. Examining again Fig.~\ref{fig:fig_1}a, one may notice that that the temporal dynamics of $\loss$ seem to slow down at lower temperatures. This is readily verified by looking at the loss auto-correlation, cf.~\cref{eq:Auto-correlation}, which shows a distinct slowing down at low $T$, as seen in Fig.~\ref{fig:loss_autocorrelation}a. To quantify this, we define the correlation half-time $\tau_{\frac12}$ as the lag time at which $\chi_\loss$ decays to $\frac{1}{2}$. Plotting $\tau_{\frac12}$ as a function of temperature, cf.~Fig.~\ref{fig:loss_autocorrelation}b, shows a clear dependence $\tau_{\frac12}\propto T^{-1}$. 

The fact that $\tau_{\frac12}$ scales as $T^{-1}$ raises three interesting insights. First, and most importantly, it is inconsistent with a picture of quadratic loss, which implies that the dynamic timescales are $T$-independent, $\tau_i=k_i^{-1}$. In contrast, we observe that $\tau_{\frac12}$ changes over 4 orders of magnitude with $T$.

Secondly, while the quadratic analogy might not hold, one may still relate the temporal timescale with the local stiffness, i.e.~$k\sim\tau^{-1}$. If this scaling relation holds, we should expect the eigenvalues of the loss Hessian to scale linearly with $T$. To test this, we measured the Hessian of the loss at 1000 randomly selected points during the exploration at steady state and calculated their eigenvalues using standard numerical procedures~\cite{pytorch, numpy}. The distribution of these eigenvalues is plotted in Fig.~\ref{fig:eigenvalues}a, clearly showing a linear scaling with $T$. 

These observations are manifestly inconsistent with a picture of an effectively quadratic loss: in the quadratic picture $\mu_\loss$ increases linearly with $T$ because the system climbs slightly higher up the confining parabolic walls, whose  stiffness is constant. Our observation suggests that the picture is quite different: $\mu_\loss$ increases in tandem with the stiffness of the confining walls, and due to a delicate balance the net result is indistinguishable from a quadratic picture, as far as static properties are considered. Below we explain this balance and show that it is related to the exponential nature of the confining walls.

Lastly, we remark that the relation $\tau_{\frac12}\sim T^{-1}$ gives rise to a distance scale $L$, defined by $L^2=T\tau_{\frac{1}{2}}$. $L$ is the distance, in parameter space, that $\theta$ would diffuse over the time $\tau_{\frac{1}{2}}$, if subject only to isotropic Gaussian noise. Since the diffusion coefficient scales with $T$, $L$ is $T$-independent. Furthermore, since $\tau_{\frac{1}{2}}$ is the correlation time, one can also interpret $L$ as a correlation length, or the distance that two nearby networks need to diffuse away from each other in order for their loss to decorrelate, i.e.~produce significantly different predictions. Since this distance scale does not depend on $T$, we conclude that it is an intrinsic property of the loss landscape, i.e.~a characteristic length scale in weight space. 

To demonstrate the effect of this length scale, we performed another numerical experiment: Starting from the minimum (the end of training phase I in Fig.~\ref{fig:loss_autocorrelation}) we let the system diffuse freely, i.e.~evolve in time according to Eq.~\eqref{eq:SDE_Langevin} but without the gradient term. This procedure samples points uniformly and isotropically around the starting point. Indeed, \cref{fig:eigenvalues}b shows that for distances smaller than $L$ the loss does note deviate significantly from its minimum value. At larger distances, $\loss$ changes by orders of magnitude over a relatively small distance.

\subsection{Summary of the numerical observations}
\begin{figure}
  \centering
  \includegraphics[width=\linewidth]{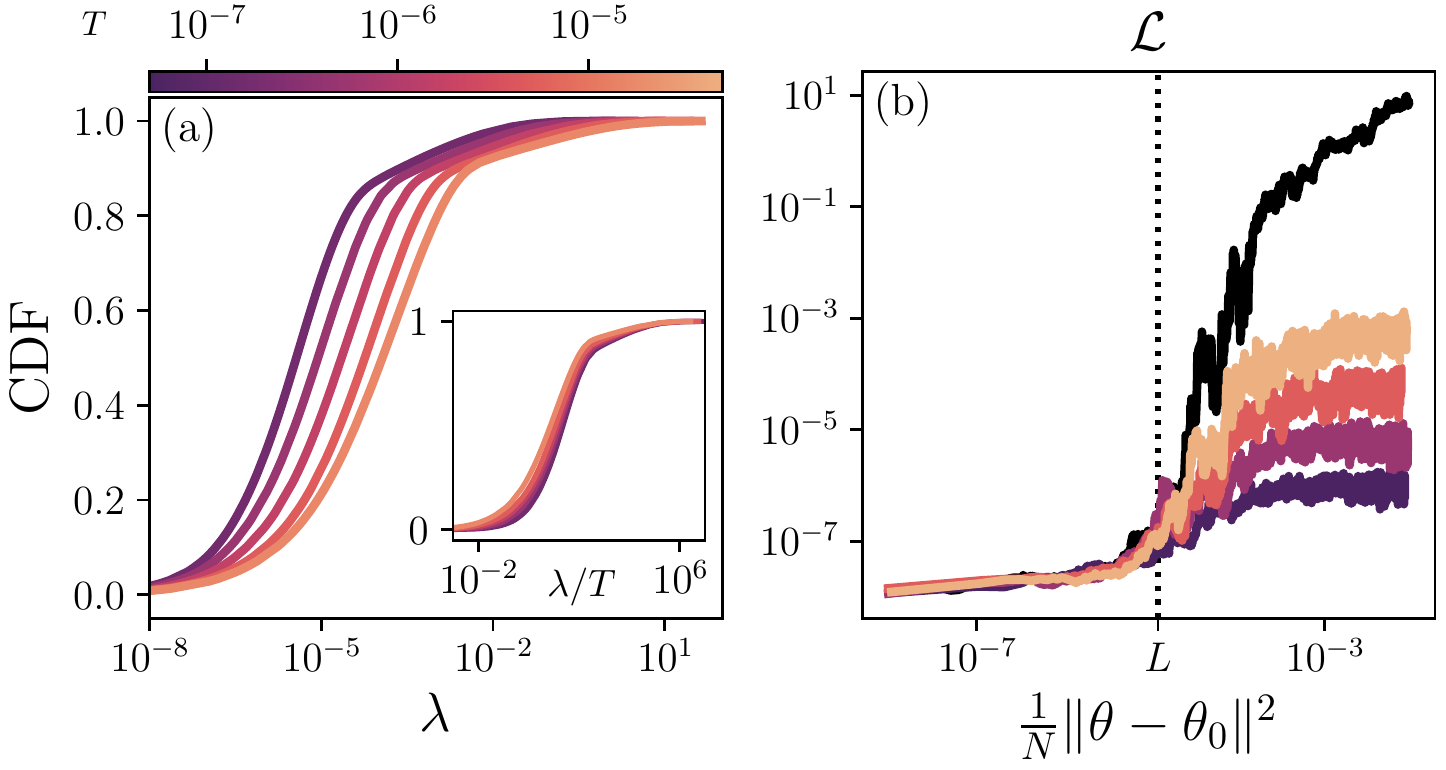}
  \caption{(a) The cumulative distribution function of the Hessian eigenvalues sampled during dynamics with $\eta=10^{-2}$, for various values of $T$. Very small negative eigenvalues are excluded from this plot.  It is seen that at higher temperatures the network explores regions with larger eigenvalues. Inset: the same data plotted as function  of $\lambda/T$ shows a collapse of the distributions, suggesting that the eigenvalues scale linearly with $T$. (b) The loss as a function of distance in weight space during the exploration. In the warm-colored curves show Langevin exploration (same color code as panel a). The black line shows the behavior in the case of pure diffusion (without gradient descent). The dashed line marks $L$, the characteristic distance in weight distance obtained from \cref{fig:loss_autocorrelation}c.
  }
  \label{fig:eigenvalues}
\end{figure}

We summarize here the main properties of the loss fluctuations in the vicinity of the minimum, described above:
\begin{enumerate}[label=(\alph*)]
  \item Both $\mu_\loss$ and $\sigma_\loss$ scale linearly with the temperature $T$, as one would expect from a quadratic loss, cf.~\cref{fig:fig_1}c.
  \item Interpreting the fluctuations as if they were generated from a quadratic loss, the effective number of degrees of freedom is found to be small and weakly $T$-dependent, cf.~\cref{fig:fig_1}d. In addition, it is closely related to the number of sample points that lie close to the decision boundary, cf.~\cref{fig:loss_autocorrelation}a.
  \item The correlation time $\tau_{1/2}$ scales as $1/T$ and the Hessian eigenvalues scale as $T$, which is inconsistent with a quadratic loss and gives rise to an emergent $T$-independent length scale $L$, cf.~\cref{fig:loss_autocorrelation} and \cref{fig:eigenvalues}.
\end{enumerate}

\section{An analytical toy model}
In order to explain our numerical observations, one need to inspect the cross entropy loss Eq.~\eqref{eq:CE_loss}. For simplicity, consider a network performing binary classification on a single training example $\{x,y\}\in \mathbb{R}\times\mathbb{R}$. Since the network is overparameterized, the networks in the low-loss region that we explore classify most of the training samples perfectly. These examples contribute negligibly to the total gradient. However, some samples lie close to the decision boundary. We focus on one such sample $\{x,y\}$ and assume without loss of generality that the correct class is $y=1$. Taking a linear approximation of $f$, the contribution of this sample to the loss is (see supplementary material for derivation)
\begin{align}
  \ell(x;\theta) &= \log\left(1+e^{f(x; \theta )}\right)\ , &
   f&= \sum_i a_i  \theta_i  + b
\end{align}
In this description, the only property of $\theta_i$ that affects the loss is its projection on $a_i$, the direction in weight-space that moves the decision boundary towards the sample point. Since all other directions in weight space are irrelevant, we ignore them and examine a one-dimensional loss function 
\begin{align} \label{eq:l1d}
 \ell_{1D}(\theta)=\log\left(1+e^{a\theta+b}\right)\approx B e^{a \theta}.
\end{align}
The approximation in \cref{eq:l1d} holds in the vicinity of the minimum because the point is well classified and  the exponent is expected to be small. We define $B\equiv e^b$, and assume for concreteness that $a>0$.

\begin{figure}[tb]
  \centering
   \includegraphics[width=0.65\linewidth]{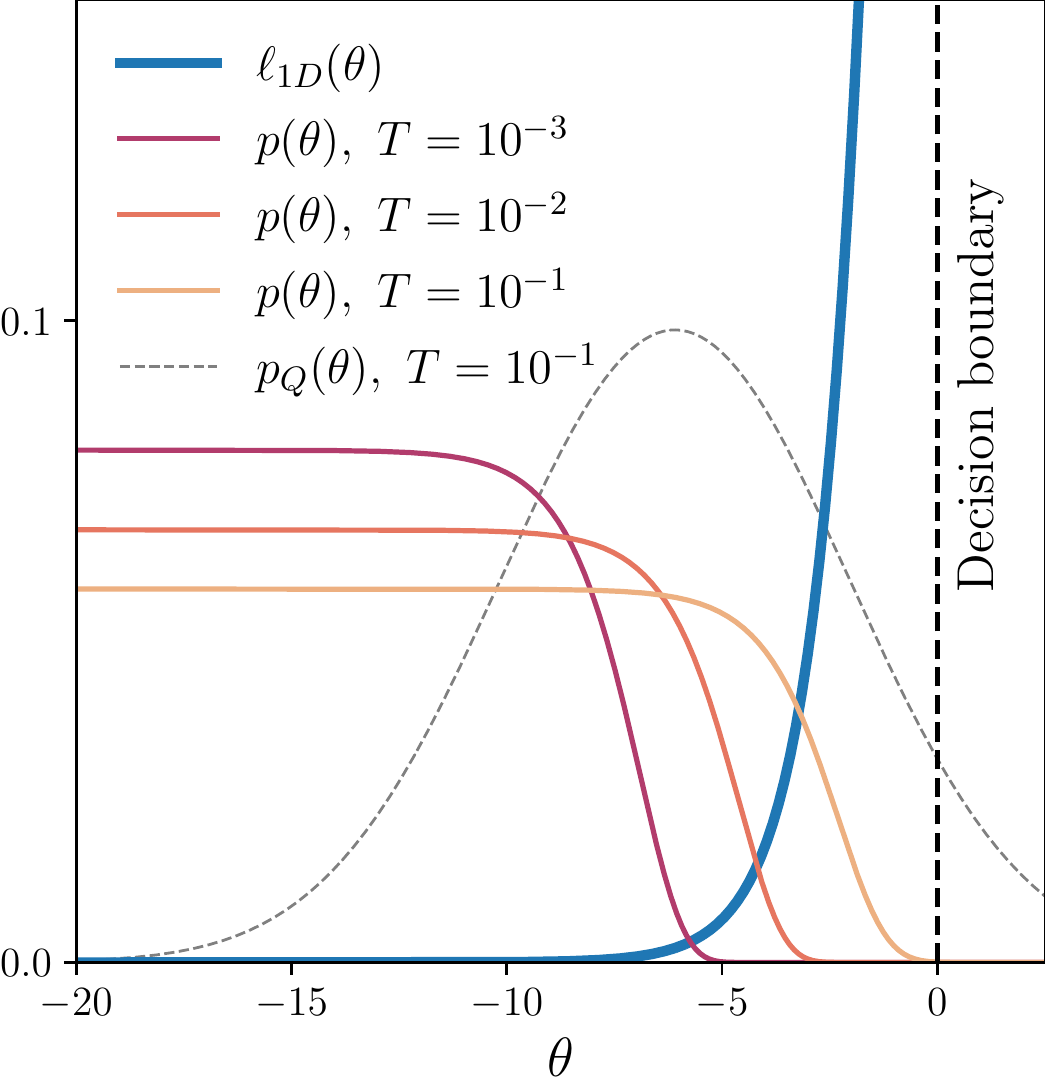}
  \caption{The loss function $\ell_{1D}$, cf.~\cref{eq:l1d}, is plotted in blue. The probability distribution of $\theta$, $p(\theta)\propto e^{-\ell_{1D}/T}$ is shown for three temperatures. It is seen that the probability distributions are qualitatively different from the probability distribution of generated by a quadratic loss, $p_Q$, which is Gaussian. For comparison, we plot $p_Q$ obtained from a quadratic approximation at $T=10^{-1}$. For this figure we chose $B=1$ and $\theta_*=20$.
  }
  \label{fig:loss_1d}
\end{figure}

The statistical mechanics of $\ell_{1D}$ can be obtained in closed form by calculating the partition function $Z(\beta)=\tfrac{1}{\theta_0}\int_{-\infty}^\infty e^{-\beta \ell_{1D}(x; \theta)} d\theta$, where $\theta_0$ is a resolution scale required to ensure the partition function is dimensionless. Formally, \cref{eq:l1d} is minimized at $\theta \to -\infty$, which effectively sets the decision boundary at infinity and prevents the integral which defines $Z$ from converging. To avoid this unphysical behavior we impose a hard cut-off at $\theta=-\theta_*$, where $\theta_*>0$, which would realistically arise when the decision boundary wanders far away and meets another sample point.

With this cutoff, the partition function $Z(\beta)$ can be obtained analytically in closed form and consequently all other ``thermodynamic'' quantities can be calculated (see supplementary information for the derivations). The main finding is that this model reproduces the properties of the loss fluctuations described above. Namely, in the limit $a \theta^*\gg1$ and $T\ll1$, both $\mu_\loss$, $\sigma_\loss$ \textit{and} the average curvature scale linearly with $T$, up to logarithmic corrections:
\begin{align}
  \begin{split}
    \mu_{\ell_{1D}} &\simeq \frac{T}{a \theta_* -\gamma +\log (T/B) }\ ,\\ 
    \sigma^2_{\ell_{1D}} & \simeq
         \frac{T^2 
         \left(a \theta _*-\gamma +\log\left({T}/{B}\right)-1 \right)}
         {\left(a \theta _* -\gamma +\log\left({T}/{B}\right)\right)^2}\ ,\\
    H_{\ell_{1D}}&=\avg{ \nabla^2_\theta \ell_{1D}} \simeq \frac{a^2 T}{a \theta_* -\gamma + \log (T/B)}\ .
    \end{split}
\end{align}
Here $\gamma\simeq 0.577$ is the Euler–Mascheroni constant. Finally, because the loss is approximately exponential in $\theta$, it features an intrinsic length scale $L \simeq a^{-1}$. We note that this length scale depends on the gradient of the network and therefore in general might differ between two different sample points that reside near the decision boundary. 

In \cref{fig:loss_1d}, we show the full loss given in \cref{eq:l1d}, and the resulting probability distribution $p(\theta)\propto e^{-\ell_{1D}/T}$ for various temperatures. It is seen that, due to the flatness of the loss, $p(\theta)$ is essentially constant at negative $\theta$ and drops sharply at the decision boundary. As $T$ grows, the probability explores regions with higher loss and, due to the exponential dependence on $\theta$, higher curvature. We compare these results against a quadratic approximation for $\ell_{1D}$, expanded around $\theta_0$ defined by $\ell_{1D}(\theta_0)=\mu_{\ell_{1D}}(T)$. It is seen that a quadratic loss is an extremely poor approximation in the low temperature limit.

\section{Summary and conclusions} To summarize our findings, we have used Langevin dynamics to investigate the geometry of the low-loss manifold of an overparameterized neural net. We find that the fluctuation statistics of the loss are a powerful probe that allows inferring geometrical insights about the loss topography. For the network studied here -- an overparameterized fully connected neural net performing a classification task on randomly distributed data -- the picture that emerges is that in the low loss region, which is explored at low temperatures, most of the sample points are well classified and do not contribute significantly to the loss. However, a small number of sample points ``pin'' the decision boundary, which fluctuates around them. At a given temperature, these fluctuations have the same statistics as fluctuations produced by a quadratic loss function, whose effective number of degrees of freedom is directly related to the number of data points constraining the decision boundary and can be immediately read off the fluctuation statistics.

However, we find that a quadratic description of the loss is fundamentally inadequate: the effective stiffness scales linearly with $T$, and correspondingly the characteristic time scale of loss fluctuations grows at low temperatures as $1/T$. These observations cannot be reconciled with a quadratic approximation of the loss. Rather, we suggest that this behavior is due to the exponential nature of the cross-entropy loss in the low $T$ regime. As we demonstrate analytically, an exponential loss function in 1D reproduces the observed fluctuation statistics in the limit of low temperature. These conclusions, of course, pertain to the simplified case studied here -- a fully connected network classifying random data. Understanding how they apply to structured data or more complicated network architectures is left for future studies.

\section{Acknowledgements}

We thank Nadav Cohen, Boaz Barak, Zohar Ringel and Stefano Recanatesi for fruitful discussions. YBS was supported by research grant ISF 1907/22 and Google Gift grant. NL would like to thank the Milner Foundation for the award of a Milner Fellowship. TK would like to acknowledge Lineage logistics for their funding. TJ was partly supported by the Raymond and Beverly Sackler Post-Doctoral Scholarship.

\bibliographystyle{apsrev4-2}
\bibliography{biblio/BiblioLangevin.bib}

\end{document}


\title{Supplementary information for:
``Charting the Topography of the Neural Network Landscape with Thermal-Like Noise''}

\author{Th\'eo Jules}
\author{Gal Brener}
\author{Tal Kachman}
\author{Noam Levi}
\author{Yohai Bar-Sinai}

\maketitle

\section{Details of the Neural Network and the dataset}
\label{app:architecture}
\paragraph{Random Dataset}

The dataset consists of 300 elements belonging to 3 classes, with inputs sampled from an isotropic multivariate normal distribution with zero mean and unit variance. The label classes are assigned randomly and uniformly. 
\paragraph{Architecture of the network}

We study a fully-connected feed-forward network with three hidden layers, of sizes [20, 20, 10], with biases.
All hidden layers are followed by rectified linear units (ReLU) activation functions. 
The network has 883 trainable weights in total, more than the number of points in the dataset.
 
\paragraph{Initialization}
\begin{figure}
  \centering
  \includegraphics[width=\linewidth]{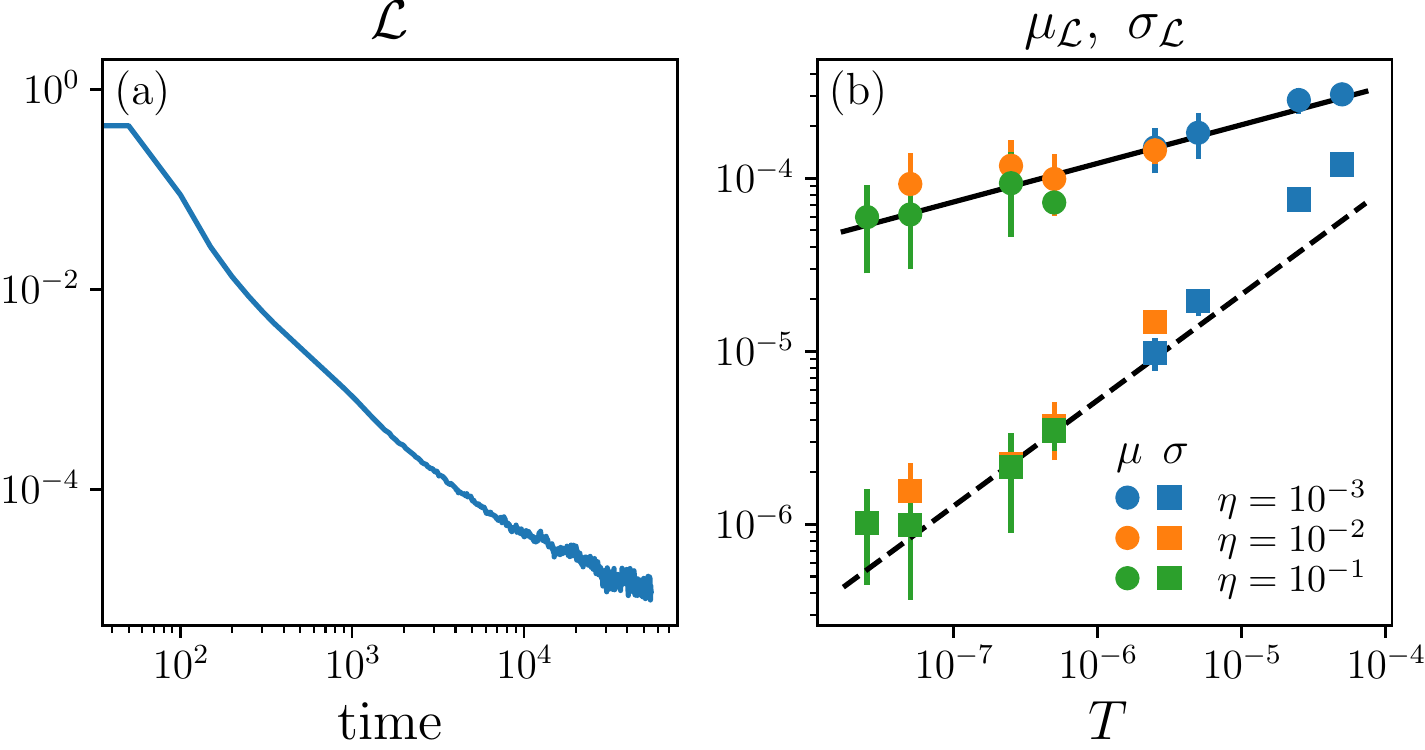}
  \caption{Overdamped Langevin dynamics following a random initialization of the network's parameters.
  (a) Evolution of the loss for $\eta = 10^{-2}$ and $T= 2.5 \times 10^{-7}$.
  (b) Dependence of the two loss moments of the loss distribution for a slice between $t=5,400$ and $t=5,600$. The solid line shows a best fit to a power law, yielding $\mu_\loss=A T^\alpha$ with $A=0.0027$ and $\alpha=0.22$. The dashed line shows the equilibrium prediction \cref{eq:S_mu_and_sigma_relation}, with no additional fitting parameters.}
  \label{fig:no_init}
\end{figure}

The neural network's weights are initialized using Xavier normal initialization, and the biases are set to zero at initialization. 

\section{Topography at initialization}
In the main text we report various statistics of the loss fluctuations during Langevin dynamics starting from a global minimum. We stress that the observations reported there are valid only if the dynamics explore the vicinity of a minimum. To demonstrate this, we show in \cref{fig:no_init}a the loss fluctuations in the case that Langevin dynamics start immediately after initialization, without the stage where a minimum is reached using Adam optimizer. It is seen that in this case the fluctuations do not converge to a time-independent steady state even after executing many more training steps.

Moreover, as shown in \cref{fig:no_init}b, if we consider the statistical properties of the loss after an extended period, where the loss is significantly small compared to initialization, the moments of the loss do not scale linearly with $T$, as they do in the vicinity of a minimum. However, $\mu_\loss$ and $\sigma_\loss$ still satisfy the equilibrium relation 
\begin{align}
  \sigma_\loss^2 = -\pd{\mu_\loss}{\beta}=T^2\pd{\mu_\loss}{T}\ .
\end{align}
For the case that $\mu_\loss=A T^\alpha$, which is observed both in the main text and in \cref{fig:no_init}, this means 
\begin{align}
  \sigma_\loss = \sqrt{\alpha AT^{\alpha+1}}\ .
  \label{eq:S_mu_and_sigma_relation}
\end{align}

\section{Temperature range}
We study the loss fluctuations at different exploration temperatures. We note that at very low temperatures, the autocorrelation time of the loss (cf.~Fig.~2b in the main text) becomes so large that meaningful sampling of the statistics would require unreasonable long training runs. We only report results for temperatures for which the autocorrelation time is at least 100 times smaller than the total run time of the simulation.

\begin{figure}
  \centering
  \includegraphics[width=\linewidth]{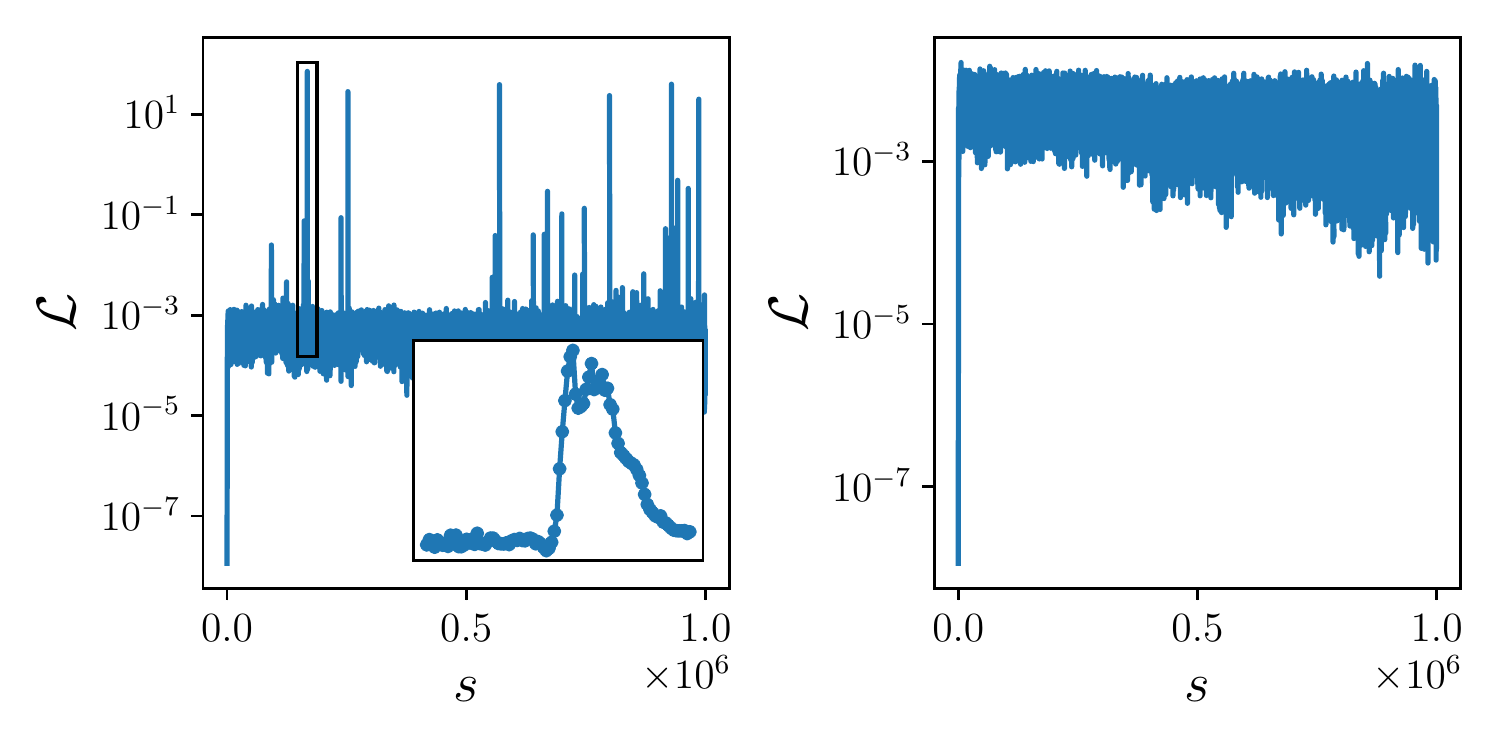}
  \caption{Loss fluctuations at high temperature. (left) Loss fluctuations for large $\eta$ ($\eta = 10^{-2}$) and high temperature ($T = 5\times 10^{-5}$). We observe catastrophic events where the loss increases and decreases by orders of magnitude in a few steps. Inset shows a blow up of one of these events (marked with a black rectangle in the main panel). (right) Drift of the loss fluctuations for $\eta = 10^{-3}$ and $T = 5\times 10^{-4}$.}
  \label{fig:High_Temp}
\end{figure}

On the other end, our temperature scale is also limited from above because at high $T$ the network starts exhibiting spiking (``catapulting'') effects during dynamics, cf.~\cref{fig:High_Temp} (left), where the loss increases and decreases by several orders of magnitude in just a few steps~\cite{Cohen2021, Lewkowycz2020, Ahn2022}. These events are short-lived, but because of the substantial loss values they significantly affect the statistics. In addition, at high $T$ the statistics are also influenced by $\eta$, which suggests that the dynamics are not thermally equilibrated and thus cannot be easily interpreted using our approach. We find that, below a certain threshold for $\eta$ (for our parameters, typically $\eta\approx 10^{-3} - 10^{-2}$), the spiking events vanish, but the overall dynamics remain unsteady even as we observe a slow drift of the mean and standard deviation of the loss fluctuations during the run, cf.~\cref{fig:High_Temp} (right).

\section{Fluctuations of a quadratic loss}
First, we consider a one-dimensional model where the loss is quadratic and can be written as
\begin{align}
\mathcal{\ell}(\theta) = \frac{k}{2} \theta^2,
\label{eq:1d_quad_loss}
\end{align}
with a stiffness $k$. In this case, $\theta$ follows a standard Ornstein-Uhlenbeck process
\begin{align}
\dot \theta(t) &= -k \theta(t) +\sqrt{2T} \xi\ ,&
\avg{\xi(t)\xi(t')}&=\delta(t-t')\ .
\end{align}

It is easily seen that the autocorrelation of $\theta$ decays exponentially with a timescale $\tau=1/k$:
\begin{align}
  \begin{split}
    \chi_{\theta} & = \sigma_\theta^{-1}\avg{\theta(t')\theta(t'+ t)}\\
    \pd{\chi_{\theta}}{t}&= \sigma_\theta^{-1}\avg{\theta(t')\dot \theta(t'+ t)}\\
    &=\sigma_\theta^{-1}\avg{
      \theta(t')\left(-k \theta(t' + t) +\sqrt{2T} \xi \right)
      }
      \\&=-k \chi_{\theta}(t)\ ,
  \end{split}
\end{align}
The autocorrelation of $\loss$ also decays exponentially with a similar timescale
\begin{align}
 \begin{split}
     \chi_\ell & = \frac{k^2}{4} \avg{\theta(t')^2\theta(t'+ t)^2}\\
     \pd{\chi_\ell}{t}&= \frac{k^2}{2}\avg{\theta(t')^2\theta(t' + t)\dot \theta(t'+ t)}\\
     &= -2k \chi_\ell
 \end{split}
\end{align}
Note that the correlation time is independent of temperature, and the curvature is constant. 

$\theta$ follows a Boltzmann distribution, which is Gaussian for a quadratic loss:
\begin{align}
  \begin{split}
    P(\theta) &= \frac{1}{Z} e^{-\beta \loss(\theta)}=\frac{1}{Z}\exp\left(-\frac{\beta k}{2} \theta^2\right)\\
    Z(\beta) &= \int e^{-\beta \loss(\theta)} \diff \theta = \sqrt{\frac{2\pi T}{k}}
  \end{split}
  \label{eq:S_distribution}
  \end{align}
where $\beta=1/T$ and $Z$ is the partition function. The loss moments can be obtained by~\cite{Kardar2007}
\begin{align}
\mu_\ell &= -\pd{\log Z}{\beta}=\frac{T}{2}\ , &
\sigma^2_\ell &= \pdd{\log Z}{\beta}=\frac{T^2}{2} \ .
\label{eq:S_moments}
\end{align}
and the full distribution is 
\begin{align}
  p(\ell)& = \frac{1}{\sqrt{\pi}} \sqrt{\frac{\beta}{\ell}}e^{-\beta \ell }\ ,
\end{align}
which is a Gamma distribution with parameters $\alpha=1/2$ and $\beta=1/T$, or alternatively a Chi-squared distribution with one degree of freedom.

For an $n$-dimensional process, when working in the diagonalizing basis, each eigendirection evolves independently of the other ones and satisfies the above equation, with $k$ replaced by the corresponding Hessian eigenvalue. The loss is a sum of contributions from all eigendirections, 
\begin{align}
  \mathcal{L} = \sum_{i=1}^{\tilde N} \ell_i \ .
\end{align}
Note that the sum runs only over the $\tilde N$ eigendirections with non-vanishing stiffness.
Since the losses of the different eigendirections are independent, the total loss is distributed as
\begin{align}
  \mathcal L \sim \Gamma\left(\alpha = \frac{\tilde N}{2}, \beta=\beta\right)
\end{align}

\section{Analytical models}

In this section, we provide analytical results for a models training with Langevin dynamics. We start by examining the classification scenario, which uses a soft-max activation followed by a cross-entropy loss, as used in the main text. We then show that for a regression task with MSE loss, using the same approximation used for the classification loss, one should expect an effectively quadratic loss.

\subsection{Logistic Regression Model}
For simplicity, we consider a neural network performing binary classification, but the calculation can be generalized for more classes.
For a given input $x$ and set of parameters $\theta$, we denote the $i$-th output of the neural network by $f_i=f(x,\theta)_i$. The predicted probability that an example $x$ belongs to class 1 is given by a softmax activation,
\begin{align}
  p_1(\theta) = \frac{\exp(f_1)}{\exp(f_1) + \exp(f_2)}
  = \frac{1}{1 + \exp(f_2-f_1)}.
\end{align}
Without loss of generality, we assume the correct class to be $1$.
Furthermore, we assume that the parameters are near the global minimum.
Consequently, $p_1 \approx 1$ and necessarily $\exp(f_2-f_1) \ll 1$.
The cross-entropy loss $\ell$ for this example reads
\begin{align}
  \ell(x,  \theta) &= -\log(p_1(\theta))= \log(1 + \exp(f_2-f_1)).
\end{align}

We further simplify the problem by linearizing $f$ (but not $\ell$) with respect to $\theta$,
\begin{align}
  f_2-f_1&= \sum_i a_i  \theta_i  + b
\end{align}
where $a=\nabla_\theta(f_2-f_1)$ and $b\in\mathbb R$. For concreteness, we take $b>0$. As written in the main text, the only property of $\theta$ that affects the loss is its projection on $a$, the direction in weight-space that moves the decision boundary towards the sample point $x$. This leads to the definition of the one-dimensional loss
\begin{align} \label{eq:S_l1d}
  \ell_{1D}(\theta)=\log\left(1+e^{a\theta+b}\right)\ .
\end{align}

The partition function of \cref{eq:S_l1d} can be calculated in closed form using hypergeometric functions. However, the expressions are significantly simplified, with negligible loss of accuracy, if we use the fact that $\exp(f_2-f_1)$ is small and thus the loss can be approximated as
\begin{align}
  \ell_{1D}(x,  \theta) &\approx B e^{a \theta} \ ,
  \label{eq:Simple_Loss}
\end{align}
with $B=e^b$.

True zero loss can only be obtained in the limit $\theta\to-\infty$. In order to obtain a physical result we impose a hard cut-off regularization at $\theta=-\theta_*$. The partition function is given by thermal averaging using the loss in \cref{eq:Simple_Loss}
\begin{align}
  \begin{split}
    \!\!\! Z 
    &=\frac{1}{\theta_0}
    \int  e^{-\beta \ell(x,\theta)} \diff \theta
    =
    \frac{1}{\theta_0}
    \int_{-\theta_*}^\infty
    \diff \theta
    \exp(-\beta B e^{a \theta  }),
  \end{split}
  \label{eq:partition}
\end{align}
where $\theta_0$ is a constant required to make the partition function dimensionless. 
\cref{eq:partition} admits a closed-form solution given by
\begin{align}
  Z = 
  -\frac{1}{a\theta_0}\text{Ei}\left(-\beta B e^{-a \theta_*}  \right)
  ,
\end{align}
where $\text{Ei}$ is the exponential integral function.

The moments of the loss distributions are easily found by differentiating the logarithm of $Z$ with respect to $\beta$, cf.~\cref{eq:S_moments}, evaluating the derivative and taking the large cut-off limit $a \theta_* \gg 1$. Concretely, we consider the regime in which $Be^{-a \theta_{*}}\ll T$, which is consistent as long as the Langevin dynamics do not push the network too far away from the minimum. In this limit, the partition function can be expanded to leading order as 
\begin{align}
  Z
  \simeq 
  \frac{\log \left(\beta  B e^{-a \theta_*
  }\right)+\gamma }{a \theta _0}-\frac{\beta  B e^{-a \theta_*
  }}{a \theta _0},
\end{align}
from which we can readily derive the first and second moments of the loss 
\begin{align}
  \avg{\ell}
  &\simeq 
  \frac{T}{a \theta _* +\log (T/B)-\gamma },
  \\ \nonumber
  \avg{\ell^2} 
  & \simeq
  \frac{T^2 
  \left(a \theta _*-\gamma +\log\left({T}/{B}\right)-1 \right)}
  {\left(a \theta _* -\gamma +\log\left({T}/{B}\right)\right)^2}\ .
\end{align}
We can further compute the average curvature eigenvalues by thermally averaging the second derivative of the loss function,
\begin{align}
  \avg{\pdd{\ell}{\theta}}
  &\simeq 
  \int 
  \pdd{\ell}{\theta} p(\theta)
  \diff \theta
  \\ \nonumber
  &=
  \frac{1}{Z}
  \int_{-\theta_0}^\infty
  \frac{ \diff \theta} { \theta_0}
  a^2
  e^{a\theta }
  \exp(-\beta B e^{a \theta}).
\end{align}
Evaluating the above integral results in
\begin{align}
  \avg{\partial_\theta^2 \ell}
  \simeq 
  \frac{a^2 T }{a \theta _*  +\log (T/B)-\gamma }.
\end{align}
Interestingly, both the loss and the Hessian scale linearly with temperature. While the first is familiar from the study of quadratic functions, common in physical systems, it is the exponential behavior of the loss which produces the $T$ dependence for the curvature.
For completeness, we also compute the entropy of the system by $S=\partial (T \log(Z))/ \partial T $, which results in the large cut-off limit to
\begin{align}
  S\simeq
  \log
  \left(
    \frac{  a \theta _*
    +\log \left( \frac{T }{B} \right)-\gamma }{a \theta _0}
    \right)
    +\frac{1}{a \theta _*
    +\log \left( \frac{T}{B} \right)
    -\gamma }\ .
  \end{align}
  This can be further expanded to
  \begin{align}
    S \simeq
    \frac{1+\log \left( \frac{T}{B} \right)-\gamma }{a \theta _*}
    +\log \left( \frac{\theta _*  }{\theta _0} \right),
  \end{align}
  demonstrating that the entropy in this model grows logarithmically with temperature.
  
  \subsection{Non-Linear Regression Model}
  We now turn to analyze the case of a network performing a regression task with MSE loss.
  Like before, the network is taken to be a multilayer perceptron~\cite{Murphy2012}, represented by the function $f(x;\theta)\!:\!\mathbb R^{d}\to \mathbb R$, which, for simplicity, we take to be smooth. 
  The network is trained on a training dataset $\{x^i, y^i\}_{i=1}^D$ where $x^i \in \mathbb R^{d}$ are the inputs, the weights are given by $\theta \in \mathbb{R}^N$ where $N$ is the number of parameters, and $y^i\in \mathbb{R}$ are the true labels. The loss is taken as the Mean Squared Error (MSE) over the entire dataset, as 
  \begin{align}
    \loss(\theta) &= \frac{1}{D} \sum_{i=1}^D \ell(x^i, y^i, \theta), &
    \ell &= \frac{1}{2}(f(x^i;\theta) - y^i)^2.
    \label{eq:MSE_loss}
  \end{align}
  Assuming a gradient based training procedure, the loss reaches its minimal value when $\nabla \loss(\theta)|_{\theta=\theta_\mathrm{opt}}=0$.
  Here, $\theta_\mathrm{opt}$ are the set of optimal weight values at the end of training. At the minimum, the gradient becomes
  \begin{align}
    \nabla \loss(\theta_\mathrm{opt})
    &= \frac{1}{D} \sum_{i=1}^D (f(x^i;\theta_\mathrm{opt}) - y^i) \nabla f(x^i;\theta_\mathrm{opt})=0 , 
    \label{eq:MSE_grad}
  \end{align}
  hence $f(x^i;\theta_\mathrm{opt}) - y^i= 0$ and we may assume that all the training samples have been fit.
  Near the minimum, the loss function can be expanded to leading order in $\theta$ as 
  \begin{align}
    \loss(\theta) 
    &\simeq 
    (\theta - \theta_\mathrm{opt})^T H(\theta_\mathrm{opt})
    (\theta - \theta_\mathrm{opt}),
    \label{eq:MSE_expand}
  \end{align}
  where 
  \begin{align}
  H\equiv \frac{1}{D} \sum_{i=1}^D \nabla_\theta f(x^i;\theta_\mathrm{opt}) \nabla_\theta^T f(x^i;\theta_\mathrm{opt}),
  \end{align}
  is a positive semi-definite Hessian matrix sans a finite number of flat directions, and we have used the fact that it is simply the product of the gradients at the minimum. 
  
  The loss in \cref{eq:MSE_expand} implies that the partition function is given by a simple gaussian integral, namely
  \begin{align}
      Z
      =\frac{1}{\theta_0^{N_c}}
      \sqrt{
      \frac{(2\pi T)^{N_c}}{\det{H}}
      },
  \end{align}
  where $N_c$ is the co-dimension of the low-loss manifold in the vicinity of the minimum and $\theta_0$ is a normalization factor. This partition function leads directly to the results for the thermally averaged loss and fluctuations quoted in Sec. 1 of the main text. 
  For completeness, we include here the also the entropy, which is given by
  \begin{align}
      S
      =
      \frac{1}{2} N_c \left[
      1+\log (2 \pi  T)
      -\log \left(\theta_0^2(\det {H})^{1/{N_c}}\right)
      \right].
  \end{align}
  Note that $S$ depends on the full eigenvalue structure of the Hessian matrix and not only on its dimensionality.

\bibliographystyle{apsrev4-2}
\bibliography{biblio/BiblioLangevin.bib}